\documentclass[11pt]{article}

\usepackage{amsmath}
\usepackage{amssymb}
\usepackage{array}
\usepackage{booktabs}
\usepackage{caption}
\usepackage{color}
\usepackage{enumitem}
\usepackage{graphicx}
\usepackage{longtable}
\usepackage[table]{xcolor}
\usepackage[dvipsnames]{xcolor}
\usepackage{inconsolata}
\usepackage{makecell}
\usepackage{multirow}
\usepackage[round]{natbib}
\usepackage{newtxtext,newtxmath}
\usepackage{pifont}
\usepackage{subcaption}
\usepackage{arydshln}
\usepackage[most]{tcolorbox}

\usepackage[T1]{fontenc}
\usepackage[a4paper, top=2.5cm, bottom=2.5cm, left=2.8cm, right=2.8cm]{geometry}
\usepackage[colorlinks=true,citecolor=RoyalBlue,linkcolor=RoyalBlue,urlcolor=RoyalBlue]{hyperref}
\usepackage[utf8]{inputenc}
\usepackage[normalem]{ulem}

\definecolor{lightbluebg}{RGB}{233,240,250}
\definecolor{blueborder}{RGB}{90,130,190}
\definecolor{lightredbg}{RGB}{250,235,235}
\definecolor{redborder}{RGB}{190,110,110}
\definecolor{lightgreenbg}{RGB}{235,247,237}
\definecolor{greenborder}{RGB}{120,170,120}

\newcolumntype{C}[1]{>{\centering\arraybackslash}p{#1}}

\newcommand{\setreturn}[1]{
  \hypertarget{return:#1}{}
}

\DeclareRobustCommand{\backtomain}[1]{%
  \hyperlink{return:#1}{\uline{Return to main text.}}%
}

\title{\textbf{From peer review nuances to best practices}}
\author{LU Sheng\thanks{Contact: \texttt{\href{mailto:arr-data@ukp.informatik.tu-darmstadt.de}{arr-data@ukp.informatik.tu-darmstadt.de}}} \\
\\
\texttt{\small \href{https://arr-data.aclweb.org/}{arr-data.aclweb.org}} \\[-0.1mm]}
\date{}

\begin{document}

\maketitle

\section{Summary}

\vspace{6pt}

\begin{tcolorbox}[
    colback=lightbluebg, colframe=blueborder,
    boxrule=0.6pt, arc=2mm, left=1.8mm, right=1.8mm, top=2mm, bottom=2mm,
    title=\textbf{Peer review nuances}
]
\begin{itemize}[leftmargin=1.2em, topsep=1.6pt, itemsep=0pt]
    \item[\ding{70}] \textbf{Paper version.} A manuscript exists in multiple versions, e.g., the \emph{initial draft} and the \emph{camera-ready} version. These versions can differ meaningfully in content.
    \item[\ding{70}] \textbf{Score version.} Multiple review score versions exist, e.g., the \emph{initial score} and the \emph{post-rebuttal score}. Different score versions correspond to different review text and paper versions.
    \item[\ding{70}] \textbf{Input format.} A paper manuscript can be fed to a model in many formats, such as \texttt{text}, \texttt{json}, \texttt{markdown}, and \texttt{image}.
\end{itemize}
\end{tcolorbox}

\vspace{8pt}

\begin{tcolorbox}[
    colback=lightredbg, colframe=redborder,
    boxrule=0.6pt, arc=2mm, left=1.8mm, right=1.8mm, top=2mm, bottom=2mm,
    title=\textbf{Why do they matter?}
]
\begin{itemize}[leftmargin=1.2em, topsep=1.6pt, itemsep=0pt]
    \item[\ding{70}] \textbf{Timely need.} Peer review research grows rapidly, making it necessary and timely to standardize how these nuances are reported and handled.
    \item[\ding{70}] \textbf{Experimental rigor.} Overlooking these nuances can undermine the validity of a study, for example, pairing pre-rebuttal review text with a post-rebuttal score.
    \item[\ding{70}] \textbf{Measurable impact.} Paper content and review scores change across versions, and the input format varies as well. These differences may affect downstream tasks.
    \item[\ding{70}] \textbf{Reproducibility.} Reporting these nuances is essential for reproducing the studies.
\end{itemize}
\end{tcolorbox}

\vspace{8pt}

\begin{tcolorbox}[
    colback=lightgreenbg, colframe=greenborder,
    boxrule=0.6pt, arc=2mm, left=1.8mm, right=1.8mm, top=2mm, bottom=2mm,
    title=\textbf{Best practices}
]
\textbf{\uline{For data providers:}}
\begin{itemize}[leftmargin=1.2em, topsep=1.6pt, itemsep=-2pt]
    \item[\ding{70}] \textbf{Cover all versions.} Preserve and release as many paper and score versions as possible.
    \item[\ding{70}] \textbf{Specify the version.} Clearly document which paper and/or score versions the dataset contains.
\end{itemize}

\vspace{6pt}

\textbf{\uline{For data users:}}
\begin{itemize}[leftmargin=1.2em, topsep=1.6pt, itemsep=-2pt]
    \item[\ding{70}] \textbf{Check the version.} Identify which paper version and score version the data contains, and verify that they are appropriate for your task.
    \item[\ding{70}] \textbf{Mind the format.} Verify whether input format matters for your setup.
    \item[\ding{70}] \textbf{Report everything.} Report the paper version, score version, and input format clearly.
\end{itemize}
\end{tcolorbox}

\clearpage

\section{Peer review nuances: version and format}

\noindent\makebox[-3.2pt][r]{\ding{70}\hspace{0.5em}}
\textbf{\uline{ARR preserves all paper and score versions.}} For ICLR 24-25 and NeurIPS 24-25, which are widely used in peer review studies, initial paper drafts and review scores may be overwritten by later revisions, and are not available from the revision history. The incomplete coverage of versions makes the data unsuitable for some tasks. For example, when training a review agent (paper as the input and human-written review as the target), using the camera-ready version may not be appropriate, as it has likely already addressed some of the issues raised in the reviews. Another example is that the review text may not be updated after the rebuttal, so a dataset may end up pairing the pre-rebuttal review text with the post-rebuttal score, creating a mismatch between review text and score version. By comparison, ARR data covers all paper versions and score versions. See Table~\ref{tab:availability}.

\begin{table}[ht]
\centering
\begin{tabular}{lcccc}
\toprule
\multirow{2.5}{*}{\textbf{venue}} & \multicolumn{2}{c}{\textbf{paper version}} & \multicolumn{2}{c}{\textbf{score version}} \\
\cmidrule(lr){2-3} \cmidrule(lr){4-5}
& \emph{initial draft} & \emph{camera-ready} & \emph{initial score} & \emph{post-rebuttal} \\
\midrule
ICLR 24-25     & \ding{55} & \ding{51} & \ding{55} & \ding{51} \\
NeurIPS 24-25  & \ding{55} & \ding{51} & \ding{55} & \ding{51} \\
\rowcolor{blue!10}
ARR             & \ding{51} & \ding{51} & \ding{51} & \ding{51} \\
\bottomrule
\end{tabular}
\caption{Availability of paper and score versions across venues. For ICLR 24-25 and NeurIPS 24-25, the initial version may be overwritten by later revisions and is thus unavailable. ARR preserves both the initial draft and the camera-ready version, as well as both the initial and post-rebuttal scores.}
\label{tab:availability}
\end{table}

\noindent\makebox[-3.2pt][r]{\ding{70}\hspace{0.5em}}
\noindent\textbf{\uline{Paper and score version are often overlooked in prior work, and input format varies widely.}} Peer review research has been growing rapidly~\citep{kuznetsov2024peerreview}, yet many data providers and data users overlook the paper version and score version, leaving them unspecified. Inspecting and specifying this information is essential for the rigor of a study (e.g., to avoid the review text and score mismatch), as well as for understanding the study setup and reproducing the results. The input format also varies across \texttt{text}, \texttt{json}, and \texttt{markdown}, and is sometimes unspecified. See Table~\ref{tab:literature}.

\begin{table}[ht]
\centering
\small
\setlength{\tabcolsep}{8pt}
\renewcommand{\arraystretch}{0.95}
\begin{subtable}{\textwidth}
\centering
\begin{tabular}{@{}llcc@{}}
\toprule
\textbf{dataset} & \textbf{source} & \textbf{paper version} & \textbf{score version} \\
\midrule
PeerRead~\citep{kang2018peerread}          & ACL, ICLR, NeurIPS        & initial       & \emph{unspecified} \\
NLP\textsc{eer}~\citep{dycke2023nlpeer}    & ARR, COLING               & multiple      & \emph{unspecified} \\
MOPRD~\citep{lin2023moprd}                 & multi-disciplinary        & multiple      & \emph{unspecified} \\
Re\textsuperscript{2}~\citep{zhang2025re2} & OpenReview conferences    & initial       & multiple \\
FMMD~\citep{zhuang2026fmmd}                & F1000Research             & multiple      & multiple \\
\bottomrule
\end{tabular}
\caption{Dataset}
\end{subtable}

\vspace{0.8em}

\begin{subtable}{\textwidth}
\centering
\begin{tabular}{@{}llccc@{}}
\toprule
\textbf{study} & \textbf{dataset} & \textbf{paper version} & \textbf{score version} & \textbf{input format}               \\ \midrule
\cite{gao2024reviewer2}       & ICLR, NeurIPS              & \emph{unspecified} & N/A                & \texttt{json}      \\
\cite{idahl2024openreviewer}  & ICLR                       & $\checkmark$       & \emph{unspecified} & \texttt{markdown}  \\
\cite{jin2024agentreview}     & ICLR                       & \emph{unspecified} & $\checkmark$       & \emph{unspecified} \\
\cite{yu2024papersea}         & ICLR, NeurIPS              & \emph{unspecified} & \emph{unspecified} & \texttt{text}      \\
\cite{gao2025reviewagents}    & ICLR, NeurIPS              & \emph{unspecified} & \emph{unspecified} & \texttt{json}      \\
\cite{sahu2025reviewertoo}    & ICLR                       & \emph{unspecified} & $\checkmark$       & \texttt{markdown}  \\
\cite{zeng2025reviewrl}       & ACL, ICLR                  & \emph{unspecified} & \emph{unspecified} & \emph{unspecified} \\
\cite{zhu2025deepreview}      & ICLR                       & \emph{unspecified} & \emph{unspecified} & \texttt{markdown}  \\
\bottomrule
\end{tabular}
\caption{Application}
\end{subtable}
\caption{The reporting of paper version, score version, and input format in prior work. \textbf{$\checkmark$} = \emph{specified}; \textbf{N/A} = \emph{not applicable}. Many data providers and data users leave the version unspecified, and the input format varies across \texttt{text}, \texttt{json}, and \texttt{markdown}.}
\label{tab:literature}
\end{table}

\clearpage

\setreturn{quantifying_differences}
\section{Quantifying version and format differences}
\label{quantifying_differences}

We conduct experiments on the EMNLP 25 dataset (2054 submissions, 3006 reviews). As an ARR venue, EMNLP 25 preserves both paper versions (\emph{initial draft} and \emph{camera-ready}) and score versions (\emph{initial} and \emph{post-rebuttal}). We study three dimensions: \textbf{paper version} (\S\ref{paper_version}), \textbf{score version} (\S\ref{score_version}), and \textbf{input format} (the format in which the paper is fed to the model, \S\ref{input_format}). We use a simple prompt that incorporates the ARR reviewer guidelines to generate reviews and scores (see Table~\ref{tab:prompt}). We experiment with gpt-oss (20B, 120B)~\citep{openai2025gptoss} and Qwen3.5 (9B, 27B)~\citep{qwen35blog}.

\subsection{Paper version}
\label{paper_version}

\begin{figure}[ht]
    \centering
    \includegraphics[width=0.7\textwidth]{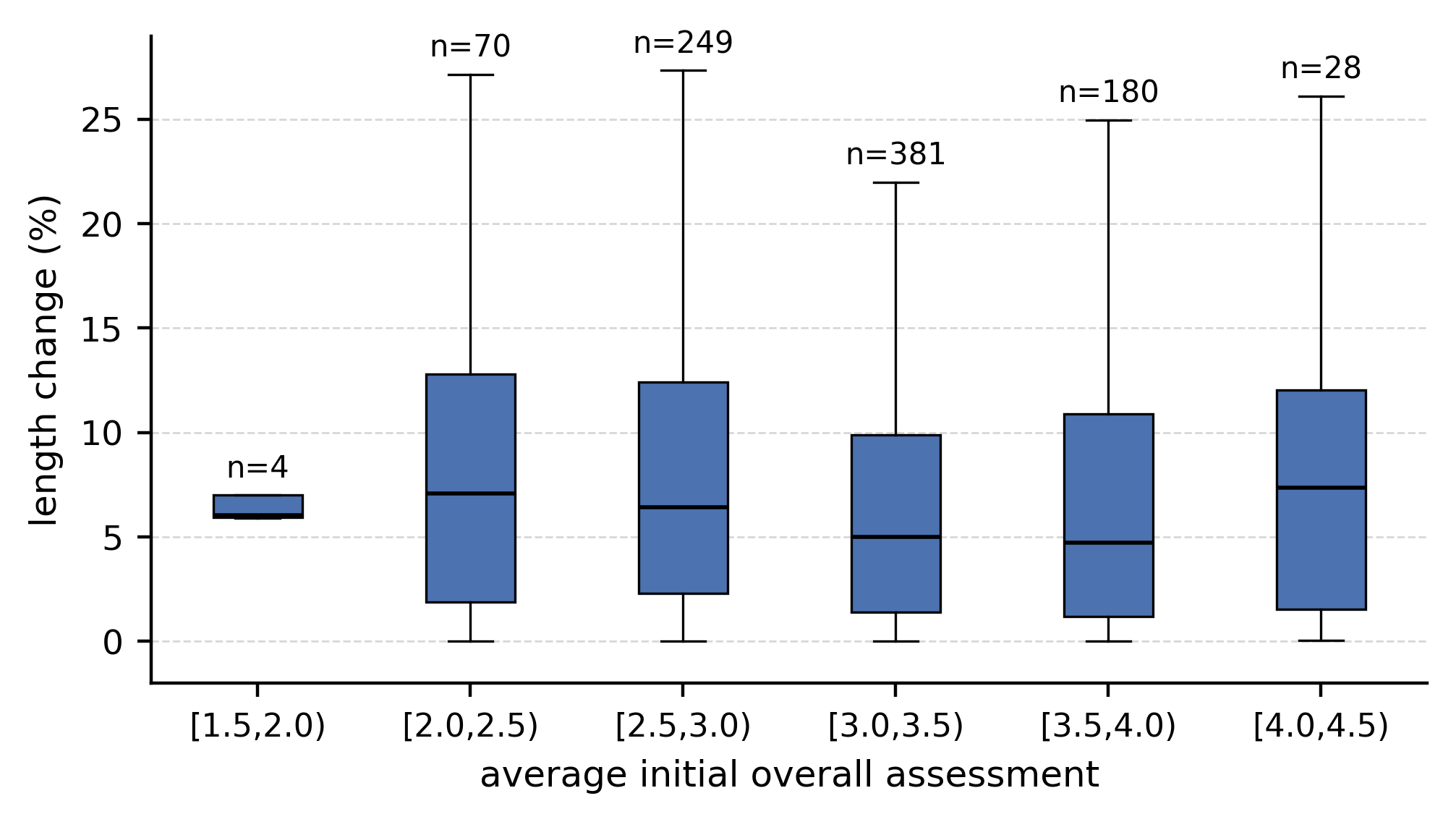}
    \caption{Distribution of the relative word-count change from the initial draft to the camera-ready version, grouped by average initial overall assessment score. Each box shows the median and interquartile range, with the number of submissions per bin ($n$). Only submissions with at least two reviews are included, so that the average of the initial scores used for binning is more reliable. The largest length changes concentrate in the $[2.0,2.5)$ and $[2.5,3.0)$ bins. The $[4.0,4.5)$ bin also shows noticeable length change, but with only $n=28$ submissions it is too small to draw a robust conclusion.}
    \label{fig:length_change_by_score}
\end{figure}

\noindent\makebox[-3.2pt][r]{\ding{70}\hspace{0.5em}}
\textbf{\uline{Submissions with mid-to-low initial scores tend to have larger revisions.}} We use the relative word-count change between the initial draft and the camera-ready version as a proxy for how much a paper was revised. Papers receiving mid-to-low initial human-assigned overall assessment scores tend to be revised more from the initial to the camera-ready version. See Figure~\ref{fig:length_change_by_score}.

\noindent\makebox[-3.2pt][r]{\ding{70}\hspace{0.5em}}
\noindent\textbf{\uline{Substance sections are revised more than Framing and Context sections.}} We quantify section-level revision using word-level normalized edit distance (wNED). We group sections into \textbf{Framing} (e.g., \emph{Introduction}, \emph{Conclusion}), \textbf{Context} (e.g., \emph{Related Work}, \emph{Motivation}), and \textbf{Substance} (e.g., \emph{Method}, \emph{Results}). Substance sections show the largest revision. See Table~\ref{tab:section_ned}.

\begin{table}[ht]
\centering
\begin{tabular}{lcccc}
\toprule
\textbf{category} & \textbf{mean} & \textbf{median} & \textbf{std.} \\
\midrule
Framing   & 0.2991 & 0.1266 & 0.3801 \\
Context   & 0.2911 & 0.1300 & 0.3768 \\
Substance & 0.4047 & 0.3073 & 0.3858 \\
\bottomrule
\end{tabular}
\caption{Word-level normalized edit distance (wNED) between the initial draft and the camera-ready version. For each submission, we group its sections into the three categories and concatenate all sections within a category, then compute one wNED per category between the versions. Statistics are aggregated over all submissions. Higher wNED indicates a larger change between versions. Substance sections show the largest revision by both mean and median wNED.}
\label{tab:section_ned}
\end{table}

\noindent\makebox[-3.2pt][r]{\ding{70}\hspace{0.5em}}
\noindent\textbf{\uline{Author information masks the paper version effect.}} We have shown that the initial draft and camera-ready version differ in content, especially for lower-scored submissions and for Substance sections. We would therefore expect the predicted overall assessment scores to differ between the two versions. A first look, however, shows almost no difference (see Table~\ref{tab:version_naive}). We find that this is because two effects push in opposite directions. Since prior work reports an authority effect where the authors or their institution may distort the evaluation of a manuscript~\citep{ye2024we,wang2026aireview}, we re-score the camera-ready version with the author information removed.\footnote{We remove the author names and affiliations shown beneath the paper title.} Comparing the initial draft against this author-information-free camera-ready version isolates the effect of paper content, and comparing the camera-ready version with and without author information isolates the author effect. Once author information is held fixed, a genuine version effect appears: the author-information-free camera-ready version is scored significantly higher than the initial draft for the larger model of each family (gpt-oss-120b and Qwen3.5-27B) (see Table~\ref{tab:version_content}). Yet unlike the commonly reported authority effect, removing author information significantly raises the predicted score for three of the four models (see Table~\ref{tab:version_author_effect}). The two effects offset each other, which is why the naive comparison looks flat.

\begin{table}[ht]
\centering
\small
\begin{subtable}{\textwidth}
\centering
\begin{tabular}{p{2cm}C{2cm}C{2cm}C{1.4cm}C{1.4cm}C{3.0cm}}
\toprule
\multirow{2.2}{*}{\textbf{model}} & \multicolumn{2}{c}{\textbf{average score}} & \multirow{2.2}{*}{\textbf{sign $p$}} & \multirow{2.2}{*}{\textbf{wilcox $p$}} & \multirow{2.2}{*}{\textbf{bootstrap 95\% CI}} \\ \cmidrule(lr){2-3}
& \textit{initial draft} & \textit{camera-ready} & & & \\ \midrule
gpt-oss-20b   & 3.84 & 3.84 & 0.8699 & 0.9957          & \raisebox{0.25ex}{[}-0.0289, 0.0266\raisebox{0.25ex}{]}          \\
gpt-oss-120b  & 3.66 & 3.70 & 0.1565 & 0.0663          & \raisebox{0.25ex}{[}-0.0035, 0.0774\raisebox{0.25ex}{]}          \\
Qwen3.5-9B    & 3.99 & 3.97 & 0.1303 & 0.1172          & \raisebox{0.25ex}{[}-0.0538, 0.0103\raisebox{0.25ex}{]}          \\
Qwen3.5-27B   & 3.81 & 3.84 & 0.0594 & \uline{0.0449} & \uline{\raisebox{0.25ex}{[}0.0046, 0.0722\raisebox{0.25ex}{]}}  \\
\bottomrule
\end{tabular}
\caption{Initial draft vs.\ camera-ready}
\label{tab:version_naive}
\end{subtable}

\vspace{8pt}

\begin{subtable}{\textwidth}
\centering
\begin{tabular}{p{2cm}C{2cm}C{2cm}C{1.4cm}C{1.4cm}C{3.0cm}}
\toprule
\multirow{2.2}{*}{\textbf{model}} & \multicolumn{2}{c}{\textbf{average score}} & \multirow{2.2}{*}{\textbf{sign $p$}} & \multirow{2.2}{*}{\textbf{wilcox $p$}} & \multirow{2.2}{*}{\textbf{bootstrap 95\% CI}} \\ \cmidrule(lr){2-3}
& \textit{initial draft} & \textit{cr w/o author} & & & \\ \midrule
gpt-oss-20b   & 3.84 & 3.85 & 0.3256          & 0.2699          & \raisebox{0.25ex}{[}-0.0151, 0.0434\raisebox{0.25ex}{]}          \\
gpt-oss-120b  & 3.66 & 3.78 & \uline{$<$0.0001} & \uline{$<$0.0001} & \uline{\raisebox{0.25ex}{[}0.0808, 0.1570\raisebox{0.25ex}{]}} \\
Qwen3.5-9B    & 3.99 & 4.01 & 0.4336          & 0.2336          & \raisebox{0.25ex}{[}-0.0114, 0.0503\raisebox{0.25ex}{]}          \\
Qwen3.5-27B   & 3.81 & 3.88 & \uline{0.0001}  & \uline{0.0001}  & \uline{\raisebox{0.25ex}{[}0.0390, 0.1067\raisebox{0.25ex}{]}} \\
\bottomrule
\end{tabular}
\caption{Initial draft vs.\ camera-ready without author information}
\label{tab:version_content}
\end{subtable}

\vspace{8pt}

\begin{subtable}{\textwidth}
\centering
\begin{tabular}{p{2cm}C{2cm}C{2cm}C{1.4cm}C{1.4cm}C{3.0cm}}
\toprule
\multirow{2.2}{*}{\textbf{model}} & \multicolumn{2}{c}{\textbf{average score}} & \multirow{2.2}{*}{\textbf{sign $p$}} & \multirow{2.2}{*}{\textbf{wilcox $p$}} & \multirow{2.2}{*}{\textbf{bootstrap 95\% CI}} \\ \cmidrule(lr){2-3}
& \textit{camera-ready} & \textit{cr w/o author} & & & \\ \midrule
gpt-oss-20b   & 3.84 & 3.85 & 0.1554          & 0.1636          & \raisebox{0.25ex}{[}-0.0104, 0.0428\raisebox{0.25ex}{]}          \\
gpt-oss-120b  & 3.70 & 3.78 & \uline{$<$0.0001} & \uline{$<$0.0001} & \uline{\raisebox{0.25ex}{[}0.0439, 0.1201\raisebox{0.25ex}{]}} \\
Qwen3.5-9B    & 3.97 & 4.01 & \uline{0.0068}  & \uline{0.0054}  & \uline{\raisebox{0.25ex}{[}0.0114, 0.0721\raisebox{0.25ex}{]}} \\
Qwen3.5-27B   & 3.84 & 3.88 & \uline{0.0313}  & \uline{0.0205}  & \uline{\raisebox{0.25ex}{[}0.0057, 0.0675\raisebox{0.25ex}{]}} \\
\bottomrule
\end{tabular}
\caption{Camera-ready vs.\ camera-ready without author information}
\label{tab:version_author_effect}
\end{subtable}
\caption{Comparison of model-predicted overall assessment scores across (a) initial draft and camera-ready, (b) initial draft and camera-ready version without author information (\textit{cr w/o author}), and (c) camera-ready version with and without author information. We report the average predicted score for each version, the $p$-values from the sign test (\textbf{sign $p$}) and the Wilcoxon signed-rank test (\textbf{wilcox $p$}), and the bootstrap 95\% confidence interval (\textbf{CI}) of the mean score difference (e.g., camera-ready $-$ initial draft). Significant results ($p<0.05$, or CI excluding 0) are \uline{underlined}. Isolating the effect of paper content shows the camera-ready version scored higher for the larger model of each family (gpt-oss-120b, Qwen3.5-27B). Removing author information significantly raises the score for three of the four models. The effects of paper content and author offset each other.}
\label{tab:version}
\end{table}

\clearpage

\subsection{Score version}
\label{score_version}

\begin{figure}[ht]
    \centering
    \begin{subfigure}[b]{0.48\textwidth}
        \centering
        \includegraphics[width=\linewidth]{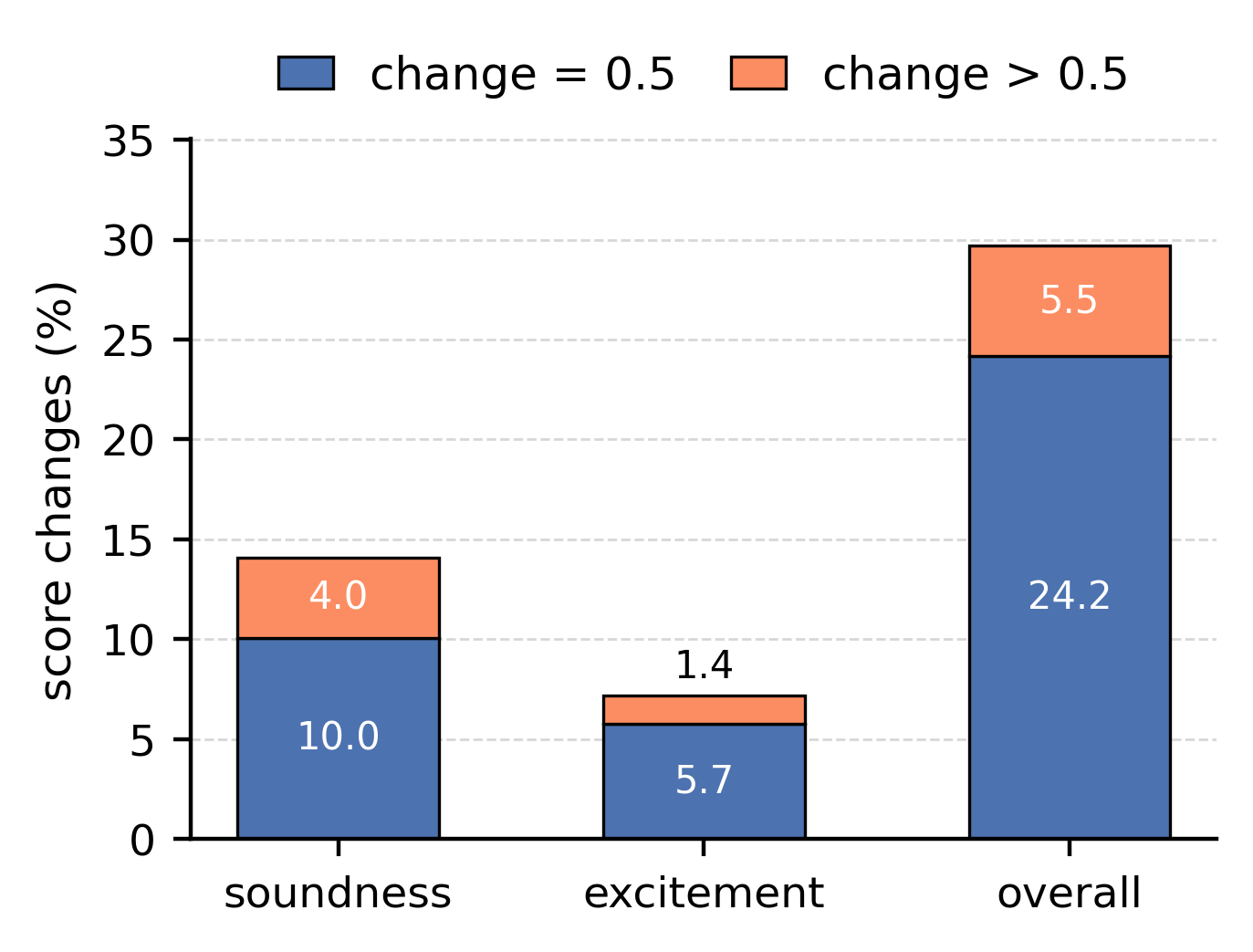}
        \caption{reviewer score changes after rebuttal}
        \label{fig:score_changes}
    \end{subfigure}
    \hfill
    \begin{subfigure}[b]{0.48\textwidth}
        \centering
        \includegraphics[width=\linewidth]{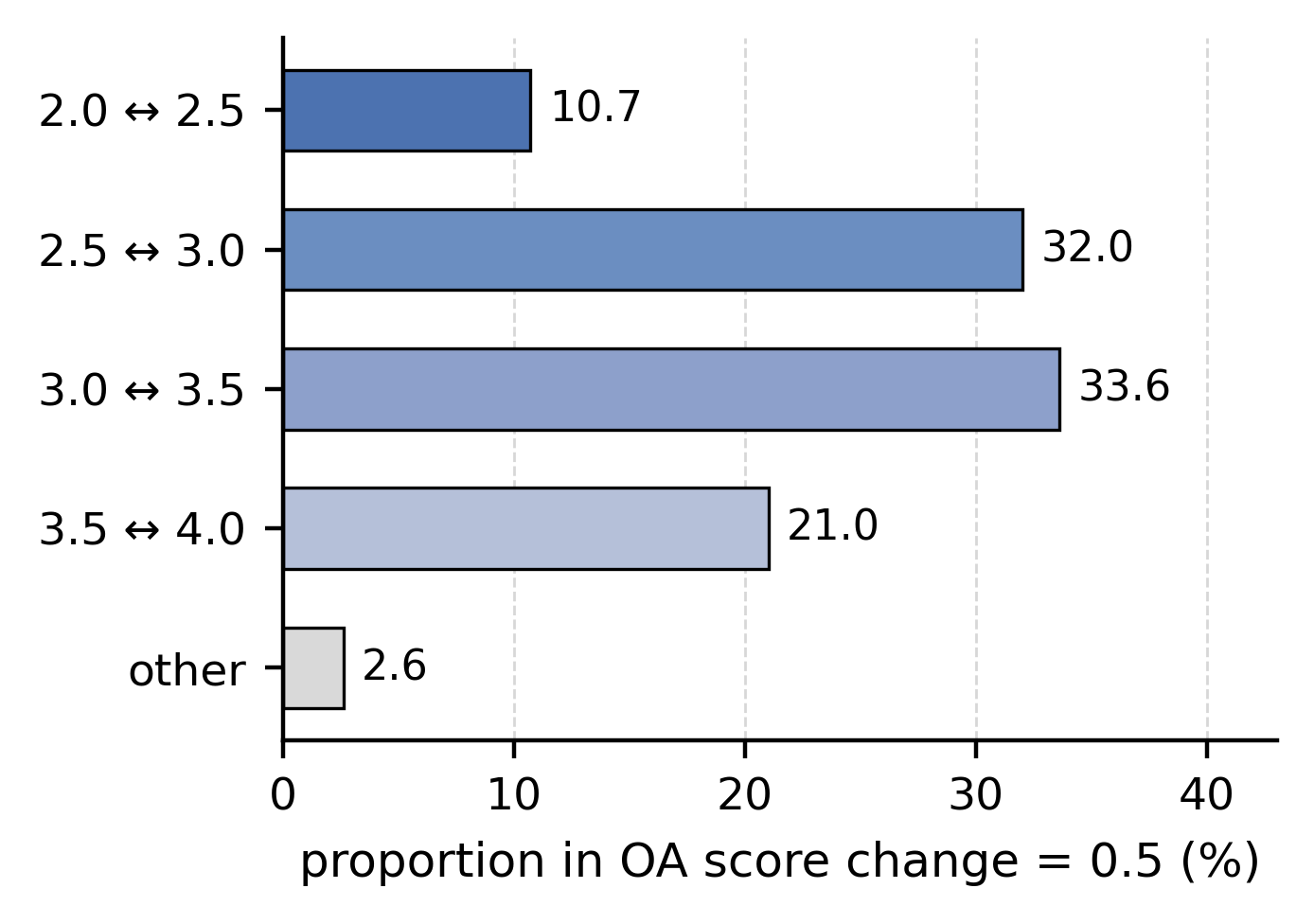}
        \caption{overall assessment change = 0.5}
        \label{fig:oa_changes_0.5}
    \end{subfigure}
    \caption{(a) Distribution of score changes, where most changes are concentrated at 0.5 points. (b) Breakdown of cases where the overall assessment change is 0.5 points, a majority occurring at \emph{decision-critical boundaries} (2.5 $\leftrightarrow$ 3.0 or 3.0 $\leftrightarrow$ 3.5).}
    \label{fig:score_changes_all}
\end{figure}

\noindent\makebox[-3.2pt][r]{\ding{70}\hspace{0.5em}}
\textbf{\uline{Nearly a third of reviews show an overall assessment score change after rebuttal.}} Across all reviews, 29.7\% show an overall assessment score change after rebuttal. The most common change is 0.5 points (24.2\%), and in 5.5\% of cases the change is greater than 0.5 points. Among the cases where the change is 0.5 points, 65.6\% occur at \emph{decision-critical boundaries} (2.5 $\leftrightarrow$ 3.0 or 3.0 $\leftrightarrow$ 3.5), where a shift may move a paper across recommendation tiers. See Figure~\ref{fig:score_changes_all}.

\setreturn{mismatch}
\noindent\makebox[-3.2pt][r]{\ding{70}\hspace{0.5em}}
\noindent\textbf{\uline{Review text and score mismatch affects model performance rankings.}} In some peer-review datasets, only the post-rebuttal score is available while the initial score is not. A user who overlooks this may pair the pre-rebuttal review text with a post-rebuttal score, creating a mismatch between the review text and the score. Here we quantify the downstream impact of this mismatch. The setup is different from that at the beginning of \S\ref{quantifying_differences}: rather than predicting from the paper, here we predict review scores from the review text (see Table~\ref{tab:score_prediction_prompt} for the prompt). We evaluate each model against either the initial or the post-rebuttal score as ground truth (the post-rebuttal score is the mismatched setting). For three of the four models, this mismatch inflates the performance. Moreover, the best performing model changes between the two settings: Qwen3.5-27B is best under the initial score, but Qwen3.5-9B overtakes it under the mismatched post-rebuttal score. See Table~\ref{tab:model_ranking}.

\begin{table}[ht]
\centering
\begin{tabular}{lccc ccc}
\toprule
\multirow{2.2}{*}{\textbf{model}} & \multicolumn{3}{c}{\textbf{$\leq$0.5$\uparrow$}} & \multicolumn{3}{c}{\textbf{rmse}$\downarrow$} \\
\cmidrule(lr){2-4} \cmidrule(lr){5-7}
 & \textit{initial score} & \textit{post-rebuttal} & $\Delta$\% & \textit{initial score} & \textit{post-rebuttal} & $\Delta$\% \\ \midrule
gpt-oss-20b       & 0.6383          & 0.7181          & $+12.50$ & 0.7824          & 0.6966          & $-10.97$ \\
gpt-oss-120b      & 0.7365          & 0.7611          & $+3.34$  & 0.6768          & 0.6441          & $-4.83$  \\ 
Qwen3.5-9B        & 0.7276          & \textbf{0.7819} & $+7.46$  & 0.6925          & \textbf{0.6353} & $-8.26$  \\
Qwen3.5-27B       & \textbf{0.7876} & 0.7455          & $-5.35$  & \textbf{0.6046} & 0.6436          & $+6.45$  \\\hdashline\noalign{\vskip 2pt}
\textit{random}   & 0.3291          & 0.3382          & $+2.77$  & 1.4667          & 1.4595          & $-0.49$  \\
\textit{majority} & 0.7094          & 0.6968          & $-1.78$  & 0.6612          & 0.6612          & $+0.00$  \\
\bottomrule
\end{tabular}
\caption{Overall assessment score prediction from review text, evaluating against the initial and post-rebuttal score as ground truth. We report the percentage of predictions within 0.5 points of the human-assigned score (\textbf{$\leq$0.5}) and the root mean squared error (\textbf{rmse}), together with the relative change ($\Delta$\%). We report the random and majority baselines. Using post-rebuttal scores leads to an inflation of performance, and the best-ranked model differs between the two score versions.}
\label{tab:model_ranking}
\end{table}

\subsection{Input format}
\label{input_format}

\begin{figure}[ht]
    \centering
    \includegraphics[width=\textwidth]{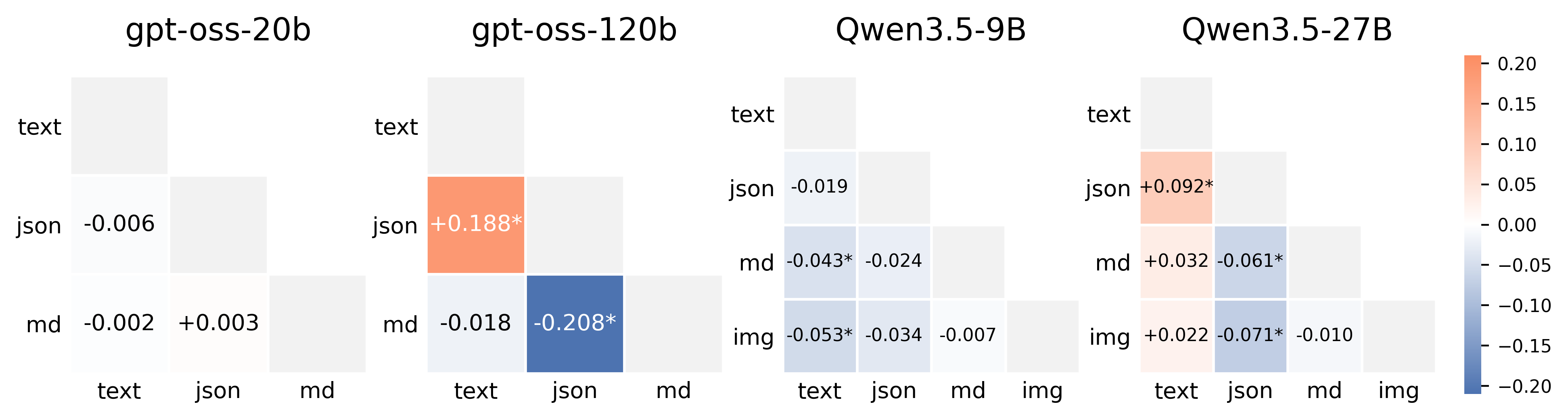}
    \caption{Differences in predicted overall assessment score across paper input formats, i.e., \texttt{text}, \texttt{json}, \texttt{markdown} (\texttt{md}), and \texttt{image} (\texttt{img}). We use the initial draft as the paper input. Each cell is the average difference between the format on the \emph{y}-axis and the format on the \emph{x}-axis (i.e., \emph{y} minus \emph{x}). The asterisk (\texttt{*}) marks differences that are statistically significant under all of the sign test, the Wilcoxon signed-rank test, and a bootstrap 95\% confidence interval excluding 0. The \texttt{image} format is evaluated only for the Qwen3.5 models, which are native vision-language models. The effect of input format on overall assessment score prediction is small for all models except gpt-oss-120b, which shows substantially larger differences. The impact of input format is mixed and model-dependent.}
    \label{fig:input_format}
\end{figure}

\noindent\makebox[-3.2pt][r]{\ding{70}\hspace{0.5em}}
\noindent\textbf{\uline{The effect of input format is mixed.}} Beyond version differences, the same paper may be fed to a model in different formats, as seen in prior work (see Table~\ref{tab:literature}). We test four formats: \textbf{\texttt{text}}, \textbf{\texttt{json}}, \textbf{\texttt{markdown}}, and \textbf{\texttt{image}}.\footnote{We convert each paper PDF to \texttt{text} and \texttt{image} using \href{https://pymupdf.readthedocs.io/en/latest/}{PyMuPDF}, and to \texttt{json} and \texttt{markdown} using \href{https://mineru.net/}{MinerU}. For \texttt{text}, elements in the original PDF are extracted as-is without any post-processing, so elements such as figures, tables, and equations are of relatively poor quality. Since \href{https://mineru.net/}{MinerU} uses a dedicated backbone model for PDF recognition, the converted \texttt{json} and \texttt{markdown} preserve these elements well. \href{https://mineru.net/}{MinerU} also extracts the embedded figures and plots, but we discard them and use only the textual part of the converted output, i.e., \texttt{json} and \texttt{markdown} are text-only.} The \texttt{image} format is tested only for the Qwen3.5 models, which are native vision-language models. Intuitively, \texttt{text} is a baseline format, as the other formats preserve more information: \texttt{json} provides more structured content, \texttt{markdown} better retains elements such as equations and tables, and \texttt{image} preserves the paper in its entirety. Our results show that the choice of input format has little effect on the predicted overall assessment scores for three of the four models. For gpt-oss-120b, however, the effect is much larger, with the \texttt{json} format yielding higher scores. The effect of input format is mixed, and should be examined on a per-model basis. See Figure~\ref{fig:input_format}.

\section{Conclusion}

We study three nuances in peer review data: paper version, score version, and input format. We characterize how the variants differ, and measure their impact on downstream tasks. Based on our findings, we offer best practices for both data providers and data users.

\vspace{8pt}

Beyond peer review research, these nuances carry broader implications for understanding the properties of large language models. Each nuance is a controlled source of variation that probes whether a model discerns the difference between variants and how it reasons over them~\citep{dycke2026counterfactual}. Our experiments already hint at such properties, for example, the mixed effect of input format. \textbf{We therefore not only call on the community to take these nuances seriously in peer review research, but also encourage the community to repurpose them as controlled probes that reveal the capabilities and limitations of large language models.}

\clearpage

\bibliography{custom}

\clearpage

\appendix

\begin{longtable}{>{\small}p{\dimexpr\textwidth-2\tabcolsep\relax}}
\phantomsection\\
\caption{The prompt used to generate reviews and scores. \backtomain{quantifying_differences}}\label{tab:prompt}\\
\toprule
\endfirsthead
\toprule
\endhead
\bottomrule
\endfoot
\bottomrule
\endlastfoot
\texttt{Given a research paper and the corresponding review guidelines, write a summary of its strengths and weaknesses. Then assign a Soundness, Excitement, and Overall Assessment score based on the summaries. Output a json dictionary.}\\
\\
\texttt{\#\# Review guidelines}\\
\\
\texttt{$\ast\ast$Summary of Strengths$\ast\ast$\newline What are the major reasons to publish this paper at a selective *ACL venue? These could include novel and useful methodology, insightful empirical results or theoretical analysis, clear organization of related literature, or any other reason why interested readers of *ACL papers may find the paper useful.}\\
\\
\texttt{$\ast\ast$Summary of Weaknesses$\ast\ast$\newline What are the concerns that you have about the paper that would cause you to favor prioritizing other high-quality papers that are also under consideration for publication? These could include concerns about correctness of the results or argumentation, limited perceived impact of the methods or findings (note that impact can be significant both in broad or in narrow sub-fields), lack of clarity in exposition, or any other reason why interested readers of *ACL papers may gain less from this paper than they would from other papers under consideration. Where possible, please number your concerns so authors may respond to them individually.}\\
\\
\texttt{$\ast\ast$Soundness$\ast\ast$\newline Given that this is a short/long paper, is it sufficiently sound and thorough? Does it clearly state scientific claims and provide adequate support for them? For experimental papers: consider the depth and/or breadth of the research questions investigated, technical soundness of experiments, methodological validity of evaluation. For position papers, surveys: consider whether the current state of the field is adequately represented and main counter-arguments acknowledged. For resource papers: consider the data collection methodology, resulting data \& the difference from existing resources are described in sufficient detail.}\\
\texttt{5 = Excellent: This study is one of the most thorough I have seen, given its type.}\\
\texttt{4.5}\\
\texttt{4 = Strong: This study provides sufficient support for all of its claims/arguments. Some extra experiments could be nice, but not essential.}\\
\texttt{3.5}\\
\texttt{3 = Acceptable: This study provides sufficient support for its major claims/arguments. Some minor points may need extra support or details.}\\
\texttt{2.5}\\
\texttt{2 = Poor: Some of the main claims/arguments are not sufficiently supported. There are major technical/methodological problems.}\\
\texttt{1.5}\\
\texttt{1 = Major Issues: This study is not yet sufficiently thorough to warrant publication or is not relevant to ACL.}\\
\\
\texttt{$\ast\ast$Excitement$\ast\ast$}\\
\texttt{How exciting is this paper for you? Excitement is subjective, and does not necessarily follow what is popular in the field. We may perceive papers as transformational/innovative/surprising, e.g. because they present conceptual breakthroughs or evidence challenging common assumptions/methods/datasets/metrics. We may be excited about the possible impact of the paper on some community (not necessarily large or our own), e.g. lowering barriers, reducing costs, enabling new applications.}\\
\texttt{We may be excited for papers that are relevant, inspiring, or useful for our own research. These factors may combine in different ways for different reviewers.}\\
\texttt{5 = Highly Exciting: I would recommend this paper to others and/or attend its presentation in a conference.}\\
\texttt{4.5}\\
\texttt{4 = Exciting: I would mention this paper to others and/or make an effort to attend its presentation in a conference.}\\
\texttt{3.5}\\
\texttt{3 = Interesting: I might mention some points of this paper to others and/or attend its presentation in a conference if there's time.}\\
\texttt{2.5}\\
\texttt{2 = Potentially Interesting: this paper does not resonate with me, but it might with others in the *ACL community.}\\
\texttt{1.5}\\
\texttt{1 = Not Exciting: this paper does not resonate with me, and I don't think it would with others in the *ACL community (e.g. it is in no way related to computational processing of language).}\\
\\
\texttt{$\ast\ast$Overall Assessment$\ast\ast$\newline If this paper was committed to an *ACL conference, do you believe it should be accepted? If you recommend conference, Findings and or even award consideration, you can still suggest minor revisions (e.g. typos, non-core missing refs, etc.). Outstanding papers should be either fascinating, controversial, surprising, impressive, or potentially field-changing. Awards will be decided based on the camera-ready version of the paper. Main vs Findings papers: the main criteria for Findings are soundness and reproducibility. Conference recommendations may also consider novelty, impact and other factors.}\\
\texttt{5 = Consider for Award: I think this paper could be considered for an outstanding paper award at an *ACL conference (up to top 2.5\% papers).}\\
\texttt{4.5 = Borderline Award}\\
\texttt{4 = Conference: I think this paper could be accepted to an *ACL conference.}\\
\texttt{3.5 = Borderline Conference}\\
\texttt{3 = Findings: I think this paper could be accepted to the Findings of the ACL.}\\
\texttt{2.5 = Borderline Findings}\\
\texttt{2 = Resubmit next cycle: I think this paper needs substantial revisions that can be completed by the next ARR cycle.}\\
\texttt{1.5 = Resubmit after next cycle: I think this paper needs substantial revisions that cannot be completed by the next ARR cycle.}\\
\texttt{1 = Do not resubmit: This paper has to be fully redone, or it is not relevant to the *ACL community (e.g. it is in no way related to computational processing of language).}\\
\\
\texttt{\#\# Output format}\\
\texttt{Output only the json dictionary and follow the json schema exactly, with no extra keys, notes, comments, or explanations:}\\
\texttt{\{``strengths'': ``...'', ``weaknesses'': ``...'', ``soundness'': ``...'', ``excitement'': ``...'', ``overall\_assessment'': ``...''\}}\\
\end{longtable}

\clearpage

\begin{longtable}{>{\small}p{\dimexpr\textwidth-2\tabcolsep\relax}}
\phantomsection\\
\caption{The prompt used to predict review scores from review text. \backtomain{mismatch}}\label{tab:score_prediction_prompt}\\
\toprule
\endfirsthead
\toprule
\endhead
\bottomrule
\endfoot
\bottomrule
\endlastfoot
\texttt{Given a paper review and the corresponding review form, recommend appropriate Soundness, Excitement, and Overall Assessment scores that the reviewer might consider giving based on their review. Provide a brief justification for each recommended score. Each justification should be concise and grounded in the review text, focusing on the main evidence for that score.}\\
\\
\texttt{\#\# Review guidelines}\\
\\
\texttt{$\ast\ast$Soundness$\ast\ast$\newline Given that this is a short/long paper, is it sufficiently sound and thorough? Does it clearly state scientific claims and provide adequate support for them? For experimental papers: consider the depth and/or breadth of the research questions investigated, technical soundness of experiments, methodological validity of evaluation. For position papers, surveys: consider whether the current state of the field is adequately represented and main counter-arguments acknowledged. For resource papers: consider the data collection methodology, resulting data \& the difference from existing resources are described in sufficient detail.}\\
\texttt{5 = Excellent: This study is one of the most thorough I have seen, given its type.}\\
\texttt{4.5}\\
\texttt{4 = Strong: This study provides sufficient support for all of its claims/arguments. Some extra experiments could be nice, but not essential.}\\
\texttt{3.5}\\
\texttt{3 = Acceptable: This study provides sufficient support for its major claims/arguments. Some minor points may need extra support or details.}\\
\texttt{2.5}\\
\texttt{2 = Poor: Some of the main claims/arguments are not sufficiently supported. There are major technical/methodological problems.}\\
\texttt{1.5}\\
\texttt{1 = Major Issues: This study is not yet sufficiently thorough to warrant publication or is not relevant to ACL.}\\
\\
\texttt{$\ast\ast$Excitement$\ast\ast$\newline How exciting is this paper for you? Excitement is subjective, and does not necessarily follow what is popular in the field. We may perceive papers as transformational/innovative/surprising, e.g. because they present conceptual breakthroughs or evidence challenging common assumptions/methods/datasets/metrics. We may be excited about the possible impact of the paper on some community (not necessarily large or our own), e.g. lowering barriers, reducing costs, enabling new applications. We may be excited for papers that are relevant, inspiring, or useful for our own research. These factors may combine in different ways for different reviewers.}\\
\texttt{5 = Highly Exciting: I would recommend this paper to others and/or attend its presentation in a conference.}\\
\texttt{4.5}\\
\texttt{4 = Exciting: I would mention this paper to others and/or make an effort to attend its presentation in a conference.}\\
\texttt{3.5}\\
\texttt{3 = Interesting: I might mention some points of this paper to others and/or attend its presentation in a conference if there's time.}\\
\texttt{2.5}\\
\texttt{2 = Potentially Interesting: this paper does not resonate with me, but it might with others in the *ACL community.}\\
\texttt{1.5}\\
\texttt{1 = Not Exciting: this paper does not resonate with me, and I don't think it would with others in the *ACL community (e.g. it is in no way related to computational processing of language).}\\
\\
\texttt{$\ast\ast$Overall Assessment$\ast\ast$\newline If this paper was committed to an *ACL conference, do you believe it should be accepted? If you recommend conference, Findings and or even award consideration, you can still suggest minor revisions (e.g. typos, non-core missing refs, etc.).}\\
\texttt{Outstanding papers should be either fascinating, controversial, surprising, impressive, or potentially field-changing. Awards will be decided based on the camera-ready version of the paper.}\\
\texttt{Main vs Findings papers: the main criteria for Findings are soundness and reproducibility. Conference recommendations may also consider novelty, impact and other factors.}\\
\texttt{5 = Consider for Award: I think this paper could be considered for an outstanding paper award at an *ACL conference (up to top 2.5\% papers).}\\
\texttt{4.5 = Borderline Award}\\
\texttt{4 = Conference: I think this paper could be accepted to an *ACL conference.}\\
\texttt{3.5 = Borderline Conference}\\
\texttt{3 = Findings: I think this paper could be accepted to the Findings of the ACL.}\\
\texttt{2.5 = Borderline Findings}\\
\texttt{2 = Resubmit next cycle: I think this paper needs substantial revisions that can be completed by the next ARR cycle.}\\
\texttt{1.5 = Resubmit after next cycle: I think this paper needs substantial revisions that cannot be completed by the next ARR cycle.}\\
\texttt{1 = Do not resubmit: This paper has to be fully redone, or it is not relevant to the *ACL community (e.g. it is in no way related to computational processing of language).}\\
\\
\texttt{\#\# Output format}\\
\texttt{Output only the json dictionary and follow the json schema exactly, with no extra keys, notes, comments, or explanations:}\\
{\raggedright\texttt{\{``soundness\_score'': ..., ``soundness\_justification'': ..., ``excitement\_score'': ..., ``excitement\_justification'': ..., ``overall\_assessment\_score'': ..., ``overall\_assessment\_justification'': ...\}}\par}
\end{longtable}

\end{document}